\documentclass{amsart}

\usepackage[T1]{fontenc}
\usepackage{textcomp} % defines \textquotedbl etc.
\usepackage{amsmath,amsfonts,amssymb,bm}
\usepackage{graphicx}
\usepackage{listings}
\usepackage[authoryear,round]{natbib}
\usepackage{booktabs}
\usepackage{empheq}
\usepackage{geometry}
\usepackage{hyperref}

\lstdefinestyle{rcompact}{
  language=R,
  basicstyle=\ttfamily\footnotesize,
  columns=fullflexible,
  keepspaces=true,
  showstringspaces=false,
  upquote=true,
  tabsize=2,
  breaklines=true,
  breakatwhitespace=true,
  numbers=left,
  numberstyle=\tiny,
  numbersep=4pt,
  aboveskip=3pt,
  belowskip=3pt,
  xleftmargin=0pt,
  frame=none
}
\newcommand{\bsy}{\boldsymbol{y}}

\newcommand{\bsbeta}{\boldsymbol{\beta}}
\newcommand{\bstheta}{\boldsymbol{\theta}}
\newcommand{\bseta}{\boldsymbol{\eta}}
\newcommand{\bseps}{\boldsymbol{\varepsilon}}
\newcommand{\bsX}{\mathbf{X}}

\newcommand{\bsZ}{\mathbf{Z}}

\newcommand{\bsR}{\mathbf{R}}
\newcommand{\bsG}{\mathbf{G}}
\newcommand{\bsV}{\mathbf{V}}
\newcommand{\bsP}{\mathbf{P}}
\newcommand{\bsM}{\mathbf{M}}
\newcommand{\bsC}{\mathbf{C}}

\newcommand{\bsT}{\mathbf{T}}
\newcommand{\bsI}{\mathbf{I}}
\newcommand{\bsr}{\mathbf{r}}

\newcommand{\RR}{\mathbb{R}}          % real line / R^n
\newcommand{\E}{\mathbb{E}}           % expectation
\newcommand{\Var}{\operatorname{Var}} % variance operator

\begin{document}

  \title[Motivating REML via Prediction-Error Covariances]{\bf Motivating REML via Prediction-Error Covariances in EM Updates for Linear Mixed Models}
  \author[Karl]{Andrew T. Karl\\
  Karl Statistical Services LLC\\
  Aurora, CO 80016, USA\\
  \texttt{akarl@asu.edu}}

\begin{abstract}
We present a computational motivation for restricted maximum likelihood (REML) estimation in linear mixed models using an expectation--maximization (EM) algorithm. At each iteration, maximum likelihood (ML) and REML solve the same mixed-model equations for the best linear unbiased estimator (BLUE) of the fixed effects and the best linear unbiased predictor (BLUP) of the random effects. They differ only in the trace adjustments used in the variance-component updates: ML uses conditional covariances of the random effects given the data, whereas REML uses prediction-error covariances from Henderson's C-matrix, reflecting uncertainty from estimating the fixed effects. Short R code makes this switch explicit, exposes the key matrices for classroom inspection, and reproduces \texttt{lme4} ML and REML fits.
\end{abstract}

  \maketitle


\newpage

\section{Introduction}
Restricted maximum likelihood (REML) estimation is often taught through likelihood algebra, error contrasts, and
projection matrices \citep[see, e.g.,][]{patterson1971,stroup2018}.
These derivations are indispensable, but they can leave students unsure what
changes computationally relative to maximum likelihood (ML) estimation (besides seeing a change in the log-likelihood function).

Rather than re-deriving REML, this note presents the iterative ML and REML updates within an expectation--maximization (EM) algorithm. The computational distinction is confined to the trace terms used to update the variance components in the M step: ML uses
covariance calculations conditional on the fixed effects, whereas REML uses
prediction-error covariances that reflect uncertainty from estimating the fixed
effects. The goal is to give students an intuition for the REML estimator that will motivate them to understand its derivation.

We work with the linear mixed model
\begin{equation}\label{eq:1}
  \bsy = \bsX\bsbeta + \bsZ\bseta + \bseps,\qquad
  \bseta \sim N(\mathbf 0,\bsG),\;
  \bseps \sim N(\mathbf 0,\bsR),
\end{equation}
where $\bsy\in\RR^n$ is the response vector, $\bsX$ is a full-rank fixed-effect model
matrix ($n\times p$), and $\bsZ$ is a random-effect model matrix ($n\times q$).
We assume $\bseta$ and $\bseps$ are independent and that $\bsG$ and $\bsR$ are
positive definite.

Section~\ref{sec:related} summarizes canonical references and the sources of the
specific ML and REML update rules used here. Section~\ref{sec:setup} introduces
Henderson's mixed-model equations and highlights the two covariance matrices that
drive the ML--REML difference. Section~\ref{sec:EM-updates} then presents the EM-type
variance-component updates side by side and isolates the role of the trace adjustments.
Appendix~\ref{app:like} records the ML and REML log-likelihoods for reference.
Appendix~\ref{app:code} provides short R code that fits Model~\eqref{eq:1} using the
EM iteration in Algorithm~1 (under simplifying assumptions on $\bsG$ and $\bsR$
in Section~\ref{sec:EM-updates}) and exposes the key matrices appearing in the update
formulas, and Appendix~\ref{sec:lme4} validates the numerical results against
\texttt{lme4}.

\section{Related work}\label{sec:related}

Canonical treatments of REML estimation of linear mixed models include
\citet{patterson1971,harville1974,searle1992}; \citet{stroup2018} provides a modern
exposition and notation closely aligned with ours. The EM framework dates to
\citet{dempster1977}; early EM algorithms for mixed models include
\citet{lairdware1982,lairdlange1987} (see \citealp{mclachlankrishnan2008} for a general
overview of EM methods). Because Algorithm~1 updates $\bsbeta$ by its generalized least squares estimator at each iteration (via Henderson's equations), the overall procedure is more precisely an expectation/conditional maximization (ECM) variant of EM \citep{mengrubin1993}.

The specific EM variance-component updates used in Section~\ref{sec:EM-updates} are
sourced as follows: the ML update forms follow \citet{karl2012,karl2013,karl2014},
while the REML trace corrections follow \citet{diffey2017}, who develop REML EM and
PX--EM algorithms using Henderson's equations \citep{henderson1959} and a
conditional REML derivation \citep{verbyla1990}. Although Newton--Raphson-type methods
are often faster in routine mixed-model fitting \citep[see, e.g.,][]{lindstrombates1988} and preferred in many popular software routines,
EM is useful here because of the form taken by the variance-component updates.

\section{Henderson's equations and covariance matrices}\label{sec:setup}

Appendix~\ref{app:like} gives the ML and REML log-likelihoods for reference. Here we
begin with Henderson's mixed-model equations, which highlight the model matrices
students see in computation
\citep{henderson1959,searle1992,mcculloch2008,diffey2017, stroup2018}.

\subsection{Mixed-model equations}

For given covariance matrices $\bsG$ and $\bsR$, define the $(p+q)\times(p+q)$ matrix
\begin{equation}\label{eq:M-matrix}
  \bsM
  =
  \begin{bmatrix}
    \bsX^\prime\bsR^{-1}\bsX & \bsX^\prime\bsR^{-1}\bsZ\\[0.25em]
    \bsZ^\prime\bsR^{-1}\bsX & \bsZ^\prime\bsR^{-1}\bsZ + \bsG^{-1}
  \end{bmatrix}
  =
  \begin{bmatrix}
    \bsM_{\beta\beta} & \bsM_{\beta\eta}\\
    \bsM_{\eta\beta}  & \bsM_{\eta\eta}
  \end{bmatrix},
\end{equation}
and its inverse (which exists under our regularity assumptions)
\begin{equation}\label{eq:C-block}
  \bsC
  = \bsM^{-1}
  =
  \begin{bmatrix}
    \bsC_{\beta\beta} & \bsC_{\beta\eta}\\
    \bsC_{\eta\beta}  & \bsC_{\eta\eta}
  \end{bmatrix}.
\end{equation}
 Solving Henderson's equations gives
\begin{equation}\label{eq:mmsol}
  \begin{bmatrix}\hat{\bsbeta}\\[0.25em]\hat{\bseta}\end{bmatrix}
  =
  \bsC
  \begin{bmatrix}
    \bsX^\prime\bsR^{-1}\bsy\\[0.25em]
    \bsZ^\prime\bsR^{-1}\bsy
  \end{bmatrix},
\end{equation}
yielding the best linear unbiased estimator (BLUE) $\hat{\bsbeta}$ and the best
linear unbiased predictor (BLUP) $\hat{\bseta}$. When estimated variance
components are plugged in, these are often called an empirical BLUE (EBLUE) and an empirical BLUP (EBLUP).

\subsection{Conditional and prediction-error covariances}\label{sec:variance-sources}

In this section and Section~\ref{sec:EM-updates}, conditional moments are evaluated
under Model~\eqref{eq:1} with $(\bsbeta,\bsG,\bsR)$ fixed at their current working
values in the EM cycle; iteration indices are suppressed. We write
$\E(\bseta\mid\bsy,\bsbeta)$ when the dependence on $\bsbeta$ is relevant. Under the
Gaussian mixed model, $\Var(\bseta\mid \bsy, \bsbeta)=(\bsZ^\prime\bsR^{-1}\bsZ+\bsG^{-1})^{-1}$
does not depend on $\bsbeta$; throughout we therefore write this matrix as
$\Var(\bseta\mid\bsy)$ (and similarly $\Var(\bsZ\bseta\mid\bsy)$).

Two covariance matrices drive the ML--REML comparison below.

\medskip
\noindent\textbf{(i) Prediction-error covariance from $\bsC$.}
For a given $\bsG$ and $\bsR$, $\bsC$ is the mean
squared error (MSE) matrix of the BLUE/BLUP; under the Gaussian model it is the
joint (sampling) covariance matrix of the BLUE and the BLUP prediction error:

\begin{equation}\label{eq:C-cov}
  \Var
    \begin{bmatrix}
      \hat{\bsbeta}\\[0.25em]
      \hat{\bseta}-\bseta
    \end{bmatrix}
  = \bsC,
\end{equation}
so $\bsC_{\beta\beta}=\Var(\hat{\bsbeta})$ and
$\bsC_{\eta\eta}=\Var(\hat{\bseta}-\bseta)$. In each EM iteration, $\bsC=\bsM^{-1}$ is evaluated at the current working values
of $(\bsG,\bsR)$, and at convergence we use $\bsC$ evaluated at the final
estimated covariance matrices as an estimated MSE (prediction-error) matrix.

Similarly, with conditional residuals
\[
  \hat{\bsr}=\bsy-\bsX\hat{\bsbeta}-\bsZ\hat{\bseta},
\]
the residual prediction error $\hat{\bsr}-\bseps$ satisfies
\[
  \Var(\hat{\bsr}-\bseps)
  =
  [\,\bsX\ \bsZ\,]\bsC[\,\bsX\ \bsZ\,]^{\prime}.
\]

\medskip
\noindent\textbf{(ii) Conditional covariance of $\bseta$ given $\bsy$.}
Under Model~\eqref{eq:1},
\[
  \Var(\bseta \mid \bsy)
  = (\bsM_{\eta\eta})^{-1}
  = \bigl(\bsZ^{\prime}\bsR^{-1}\bsZ+\bsG^{-1}\bigr)^{-1},
\]
This conditional covariance depends only on
$\bsZ$, $\bsG$, and $\bsR$, whereas $\bsC_{\eta\eta}=\Var(\hat{\bseta}-\bseta)$
also reflects the additional variability induced by estimating $\bsbeta$ (through
the full inverse $\bsC=\bsM^{-1}$). $\bsC_{\eta\eta}$ is the inverse Schur complement of the fixed-effect block
$\bsM_{\beta\beta}$:
\begin{equation}\label{eq:schur-Cetaeta}
  \bsC_{\eta\eta}
  = \bigl(\bsM_{\eta\eta} - \bsM_{\eta\beta}\bsM_{\beta\beta}^{-1}\bsM_{\beta\eta}\bigr)^{-1}.
\end{equation}
Identity \eqref{eq:schur-Cetaeta} makes explicit how the fixed-effect
matrix $\bsX$ enters the prediction-error covariance through the cross-blocks of
$\bsM$.

With these two variance sources in hand, we can now write the ML and REML EM
updates side by side.

\section{EM updates: ML vs.\ REML}\label{sec:EM-updates}

The EM updates in this section follow \citet{karl2012,karl2013,karl2014} and
\citet{diffey2017}; derivations are omitted to keep the focus on the update rules. Starting from initial $(\tau^2,\sigma^2)$, we alternate E and M steps until the
relative change in the variance components is below a tolerance. For readability we
suppress iteration indices, so the hats in \eqref{eq:gammag}--\eqref{eq:sig-REML}
denote the updated values computed from the current estimates.

\subsection{E step}

Given current variance components $(\bsG,\bsR)$, the E step uses the conditional
distribution of the random effects, $\bseta\mid\bsy,\bsbeta$, which is multivariate
normal under Model~\eqref{eq:1}. Its conditional mean is
\[
  \hat{\bseta}
  = \E(\bseta\mid\bsy,\bsbeta)
  = (\bsM_{\eta\eta})^{-1}\bsZ^\prime\bsR^{-1}(\bsy-\bsX\bsbeta),
\]
and its conditional covariance is
\[
  \Var(\bseta\mid \bsy) = (\bsM_{\eta\eta})^{-1}.
\]
In the R implementation in Appendix~\ref{app:code}, we obtain $\hat{\bsbeta}$ and the
corresponding BLUP $\hat{\bseta}$ simultaneously by solving Henderson's mixed-model
equations~\eqref{eq:mmsol}. In this sense Algorithm~1 is an EM-type iteration with a
conditional maximization step for $\bsbeta$ \citep{mengrubin1993}. The E step calculations
are the same for ML and REML; the criteria differ only in the M step for the variance components.

\subsection{M step: variance components}
To obtain closed form variance component updates and avoid extraneous notation, we specialize in this subsection to the common case
$\bsG=\tau^2\bsI_q$ and $\bsR=\sigma^2\bsI_n$. Corresponding updates for block-diagonal structures, with an analogous interpretation, are given by
\citet{karl2013} and \citet{diffey2017}.

\medskip
\noindent\textbf{Random-effect variance.}
For $\bsG=\tau^2\bsI_q$, the EM updates are
\begin{empheq}[box=\fbox]{align}
  \hat{\tau}^2_{\mathrm{ML}}
    &= \frac{\hat{\bseta}^\prime \hat{\bseta}}{q}
       + \frac{1}{q}\,\operatorname{tr}\!\left\{(\bsM_{\eta\eta})^{-1}\right\} \notag\\
    &= \frac{\hat{\bseta}^\prime \hat{\bseta}}{q}
       + \frac{1}{q}\,\operatorname{tr}\!\left\{\Var(\bseta\mid\bsy)\right\}, \label{eq:gammag}\\[2pt]
  \hat{\tau}^2_{\mathrm{REML}}
    &= \frac{\hat{\bseta}^\prime \hat{\bseta}}{q}
       + \frac{1}{q}\,\operatorname{tr}\!\left\{\bsC_{\eta\eta}\right\} \notag\\
    &= \frac{\hat{\bseta}^\prime \hat{\bseta}}{q}
       + \frac{1}{q}\,\operatorname{tr}\!\left\{\Var(\hat{\bseta}-\bseta)\right\}. \label{eq:gammag-REML}
\end{empheq}
For ML, the update follows from the conditional second moment under the Gaussian model,
\[
\E(\bseta\bseta^\prime \mid \bsy,\bsbeta)
  = \hat{\bseta}\hat{\bseta}^\prime + \Var(\bseta\mid\bsy),
\]
so that $\E(\bseta^\prime\bseta\mid\bsy,\bsbeta)=\hat{\bseta}^\prime\hat{\bseta}
+\operatorname{tr}\{\Var(\bseta\mid\bsy)\}$.
For REML, the variance adjustment uses the BLUP prediction-error covariance
$\bsC_{\eta\eta}=\Var(\hat{\bseta}-\bseta)$.
This differs from the ML adjustment $(\bsM_{\eta\eta})^{-1}=\Var(\bseta\mid \bsy)$ because it also reflects uncertainty from estimating $\bsbeta$; see the Schur-complement identity \eqref{eq:schur-Cetaeta}.

\medskip
\noindent\textbf{Residual variance.}
For $\bsR=\sigma^2\bsI_n$, the EM updates are
\begin{empheq}[box=\fbox]{align}
  \hat{\sigma}^2_{\mathrm{ML}}
    &= \frac{\hat{\bsr}^{\prime}\hat{\bsr}}{n}
       + \frac{1}{n}\,\operatorname{tr}\!\left\{\bsZ\,(\bsM_{\eta\eta})^{-1}\,\bsZ^{\prime}\right\} \notag\\
    &= \frac{\hat{\bsr}^{\prime}\hat{\bsr}}{n}
       + \frac{1}{n}\,\operatorname{tr}\!\left\{\Var(\bsZ\bseta \mid \bsy)\right\} \;, \label{eq:sigM}\\[2pt]
  \hat{\sigma}^2_{\mathrm{REML}}
    &= \frac{\hat{\bsr}^{\prime}\hat{\bsr}}{n}
       + \frac{1}{n}\,\operatorname{tr}\!\left\{
             [\,\bsX\ \bsZ\,]\bsC[\,\bsX\ \bsZ\,]^{\prime}
           \right\} \notag\\
    &= \frac{\hat{\bsr}^{\prime}\hat{\bsr}}{n}
       + \frac{1}{n}\,\operatorname{tr}\!\left\{
            \Var(\hat{\bsr} - \bseps)
           \right\}\;.\label{eq:sig-REML}
\end{empheq}

The first term $\hat{\bsr}'\hat{\bsr}/n$ is the average squared conditional residual, and the trace term is a variance adjustment. Thus, as in \eqref{eq:gammag}--\eqref{eq:gammag-REML}, ML and REML share the same average squared term and differ only in whether the variance adjustment is conditional (ML) or prediction-error-based (REML).

Algorithm~1 provides a concise summary of the EM algorithm for this model, with the REML/ML choice contained in Step~4.

\begin{center}
\setlength{\fboxsep}{8pt}
\fbox{%
\begin{minipage}{0.95\linewidth}
{
\textbf{Algorithm 1 (EM iteration for the simple LMM; ML vs.\ REML differs only in trace terms).}
\begin{enumerate}
\item Initialize $(\tau^2,\sigma^2)$.
\item For current $(\tau^2,\sigma^2)$, set $\bsG=\tau^2\bsI_q$, $\bsR=\sigma^2\bsI_n$ and form
      $\bsG^{-1}$ and $\bsR^{-1}$. Build $\bsM$ in \eqref{eq:M-matrix} and compute $\bsC=\bsM^{-1}$.
      Also compute $(\bsM_{\eta\eta})^{-1}$.
\item Solve Henderson's equations \eqref{eq:mmsol} to obtain $(\hat{\bsbeta},\hat{\bseta})$ and set
      $\hat{\bsr}=\bsy-\bsX\hat{\bsbeta}-\bsZ\hat{\bseta}$.
\item Define the trace-adjustment matrices $\bsT_\tau$ and $\bsT_\sigma$.
      \emph{ML:} set $\bsT_\tau=(\bsM_{\eta\eta})^{-1}$ and
      $\bsT_\sigma=\bsZ(\bsM_{\eta\eta})^{-1}\bsZ^\prime$.
      \emph{REML:} set $\bsT_\tau=\bsC_{\eta\eta}$ and
      $\bsT_\sigma=[\,\bsX\ \bsZ\,]\bsC[\,\bsX\ \bsZ\,]^\prime$.
\item Update
\[
  \sigma^2 \leftarrow \frac{\hat{\bsr}^\prime\hat{\bsr} + \operatorname{tr}(\bsT_\sigma)}{n},
  \qquad
  \tau^2 \leftarrow \frac{\hat{\bseta}^\prime\hat{\bseta} + \operatorname{tr}(\bsT_\tau)}{q}.
\]
\item Repeat Steps 2--5 until the relative change in $(\tau^2,\sigma^2)$ is below a tolerance.
\end{enumerate}
}
\end{minipage}}
\end{center}

\paragraph{Connection to the linear-model REML estimator.}
Setting $\bsZ=\mathbf 0$ reduces the mixed model to $\bsy=\bsX\bsbeta+\bseps$; in this
ordinary least squares case, we would not run EM, and we make the substitution only to
show that the REML update \eqref{eq:sig-REML} reproduces the familiar REML estimator (for students skeptical of the $\frac{1}{n}$ factor).
Then the trace term in \eqref{eq:sig-REML} becomes
$\operatorname{tr}\{\sigma^2\bsX(\bsX'\bsX)^{-1}\bsX'\}=\sigma^2\operatorname{tr}(\bsP_X)=\sigma^2 p$, where $\bsP_X=\bsX(\bsX^\prime\bsX)^{-1}\bsX^\prime$ is the usual hat matrix,
and $\hat{\bsbeta}$ (hence $\hat{\bsr}=\bsy-\bsX\hat{\bsbeta}$) is invariant to $\sigma^2$.
Thus \eqref{eq:sig-REML} yields
\[
\sigma_{t+1}^{2}=\frac{\hat{\bsr}'\hat{\bsr}}{n}+\frac{p}{n}\sigma_{t}^{2}.
\]
Since $p<n$, the iteration converges to the fixed point
$\sigma_{\infty}^{2}=\hat{\bsr}'\hat{\bsr}/(n-p)$.

\section{Closing remarks}
The boxed updates \eqref{eq:gammag}--\eqref{eq:sig-REML} and
Table~\ref{tab:ml-reml-at-a-glance} summarize the computational distinction in this
EM presentation. At a fixed set of variance components, ML and REML solve the
same mixed-model equations for $(\hat{\bsbeta},\hat{\bseta})$; the difference is
confined to which covariance matrix supplies the variance-adjustment (trace) terms
in the variance-component updates.

As a classroom entry point, a teacher can implement the ML updates and then swap in the
REML trace corrections to see what changes (inside a single iteration, recalling that the algorithm must run to convergence). Readers who want to see why these are
the appropriate adjustments---and how they arise from the REML criterion---should
consult standard derivations of REML and EM updates; see \citet{karl2013,diffey2017,stroup2018}
and the references therein.

Although we specialized to $\bsG=\tau^2\bsI_q$ and $\bsR=\sigma^2\bsI_n$ to keep the variance-component M step algebra simple, the underlying distinction persists for more general covariance structures: ML-type EM updates use conditional second moments (and hence conditional covariances) of the random effects given the observed data \citep[e.g.,][]{karl2014}, whereas REML-type EM updates replace these with prediction-error covariances that account for estimating the fixed effects \citep[e.g.,][]{diffey2017}.

\begin{table}[t]
\centering
\caption{Trace terms in the EM variance-component updates under ML and REML, where $\bsC=\bsM^{-1}$}
\label{tab:ml-reml-at-a-glance}
\begin{tabular}{@{}lll@{}}
\toprule
Update & ML trace adjustment uses & REML trace adjustment uses\\
\midrule
$\tau^2$ in \eqref{eq:gammag}--\eqref{eq:gammag-REML}
& $\operatorname{tr}\!\{(\bsM_{\eta\eta})^{-1}\}/q$
& $\operatorname{tr}\!\{\bsC_{\eta\eta}\}/q$\\
$\sigma^2$ in \eqref{eq:sigM}--\eqref{eq:sig-REML}
& $\operatorname{tr}\!\{\bsZ(\bsM_{\eta\eta})^{-1}\bsZ^\prime\}/n$
& $\operatorname{tr}\!\{[\,\bsX\ \bsZ\,]\bsC[\,\bsX\ \bsZ\,]^\prime\}/n$\\
\bottomrule
\end{tabular}
\end{table}

For further reading and additional code, see the internal functions of the \texttt{JM},
\texttt{mvglmmRank}, \texttt{RealVAMS}, and \texttt{GPvam} packages
\citep{riz09,karl2014,mvglmmRank2023,gpvam2024,realvams2024,broatch2018} for EM-type
implementations in more complex models, including richer structures for $\bsG$
and $\bsR$ (in the presence of missing data) and multiple non-normal responses.

\section*{Disclosure Statement}

The author reports there are no competing interests to declare.

\section*{Data Availability Statement}

No new data were generated for this note. The R code to reproduce the numerical results
(including Table~\ref{tab:lme4-validation}) is provided as supplementary material with the submission.

\section*{Acknowledgments}
The author used ChatGPT (GPT-5.2 Pro) to assist with proofreading and rewording. It was also used to help with LaTeX formatting and to structure and comment the R code. All AI-suggested changes were reviewed and verified by the author, who remains responsible for the content.

\appendix

\section{Log-Likelihood Functions}\label{app:like}

Under Model~\eqref{eq:1} we have
\[
  \bsy \sim N\bigl(\bsX\bsbeta,\bsV(\bstheta)\bigr),
  \qquad
  \bsV(\bstheta) = \bsZ\bsG(\bstheta)\bsZ^\prime + \bsR(\bstheta),
\]
where $\bstheta$ collects the variance component parameters indexing $\bsG(\bstheta)$ and
$\bsR(\bstheta)$. The ML log-likelihood is
\[
  \ell_{\mathrm{ML}}(\bsbeta,\bstheta)
  = -\tfrac12\Bigl\{
      \log|\bsV|
      + (\bsy-\bsX\bsbeta)^\prime\bsV^{-1}(\bsy-\bsX\bsbeta)
      + n\log(2\pi)
    \Bigr\},
\]
and the REML log-likelihood \citep{patterson1971,harville1974,verbyla1990,
searle1992,stroup2018} is
\[
  \ell_{\mathrm{REML}}(\bstheta)
  = -\tfrac12\Bigl\{
      \log|\bsV|
      + \log|\bsX^\prime\bsV^{-1}\bsX|
      + \bsy^\prime \bsP \bsy
      + (n-p)\log(2\pi)
    \Bigr\},
\]
with
\[
  \bsP
  = \bsV^{-1}
    - \bsV^{-1}\bsX(\bsX^\prime\bsV^{-1}\bsX)^{-1}\bsX^\prime\bsV^{-1}.
\]

\section{EM code for teaching}\label{app:code}
The function \texttt{em\_lmm()} below fits the linear mixed model with
$\bsG=\tau^2\bsI_q$ and $\bsR=\sigma^2\bsI_n$. It is adapted from the
\texttt{N\_mov} function in the CRAN package \texttt{mvglmmRank} \citep{mvglmmRank2023}, but is written to
make the matrices in the update formulas explicit by forming $\bsM$ and
$\bsC=\bsM^{-1}$ at each iteration.

Inputs are compatible model objects $(\bsy,\bsX,\bsZ)$.  
The commented lines above
the function show how to extract these from an example \texttt{mvglmmRank} fit for modeling NFL scores with a multiple membership $\bsZ$ structure \citep{harville77};
Appendix~\ref{sec:lme4} shows the analogous extraction from an \texttt{lme4}
\texttt{lmer()} fit. (In sports-ranking applications such as \texttt{mvglmmRank}, the fixed-effect home-field advantage estimator $\hat{\bsbeta}$ can be biased under unbalanced or nonrandom schedules, $\bsZ$; see \citet{karl2021}.) For clarity, the implementation uses dense matrix
operations; large problems should use sparse linear algebra. To keep the listing
short, we do not evaluate the (RE)ML log-likelihood inside \texttt{em\_lmm()}; however,
the supplemental validation script computes the final log-likelihood and matches
\texttt{lme4}. Note that REML and ML log-likelihoods are defined on different likelihood scales (restricted vs.\ full) and should not be compared directly across likelihood types.

\medskip
{
\begin{lstlisting}
# Example: extract y, X, Z from an nfl2012 fit
# Uncomment these lines to load example data from mvglmmRank
# library(mvglmmRank)
# data(nfl2012)
# res <- mvglmmRank(nfl2012, method = "N.mov", verbose = FALSE, tol.n=1e-16)
# res$parameters #for comparison with em_lmm output
# Z   <- res$N.output$Z
# X   <- res$N.output$X
# y   <- res$N.output$Y

em_lmm <- function(y, X, Z, REML = FALSE,
                   maxit = 100, tol = 1e-7,
                   tau2_init = 1, sigma2_init = 1) {
  
  # LMM: y = X beta + Z eta + eps; eta ~ N(0, tau^2 I), eps ~ N(0, sigma^2 I).
  # REML toggles which covariance appears in the variance-component trace adjustments.
  
  y  <- as.matrix(y)
  X  <- as.matrix(X)
  Z  <- as.matrix(Z)
  XZ <- cbind(X, Z)
  
  n <- nrow(X)
  p <- ncol(X)
  q <- ncol(Z)
  if (nrow(Z) != n) stop("X and Z must have the same number of rows.")
  if (qr(X)$rank < ncol(X)) stop("X must be full rank.")
  
  # Initial values for variance components
  tau2   <- tau2_init
  sigma2 <- sigma2_init
  
  # Initial values for beta and eta
  beta <- matrix(0, p, 1)
  eta  <- matrix(0, q, 1)
  
  for (iter in 1:maxit) {
    
    ## --- Henderson solve: compute BLUE/BLUP and needed covariance blocks ---    
    # G^{-1} and R^{-1}
    G_inv <- diag(1 / tau2, q)    # tau^{-2} I_q
    R_inv <- diag(1 / sigma2, n)  # sigma^{-2} I_n
    
    XtRinv <- t(X) %*% R_inv      # X' R^{-1}
    ZtRinv <- t(Z) %*% R_inv      # Z' R^{-1}
    
    # Blocks of M
    M_betabeta <- XtRinv %*% X
    M_betaeta  <- XtRinv %*% Z
    M_etabeta  <- t(M_betaeta)
    M_etaeta   <- ZtRinv %*% Z + G_inv   # Z' R^{-1} Z + G^{-1}
    
    M <- rbind(
      cbind(M_betabeta, M_betaeta),
      cbind(M_etabeta,  M_etaeta)
    )
    
    # Right-hand side [X' R^{-1} y; Z' R^{-1} y]
    rhs <- rbind(
      XtRinv %*% y,
      ZtRinv %*% y
    )
    
    # Henderson's mixed model equations:
    # [beta_hat; eta_hat] = C [X' R^{-1} y; Z' R^{-1} y],
    # where C = M^{-1} is the "C-matrix"
    C   <- chol2inv(chol(M))
    sol <- C %*% rhs
    
    beta <- sol[1:p, , drop = FALSE]
    eta  <- sol[(p + 1):(p + q), , drop = FALSE]
    
    # C_{eta,eta}: prediction-error covariance for eta, Var(eta_hat - eta)
    C_etaeta <- C[(p + 1):(p + q), (p + 1):(p + q), drop = FALSE]
    
    # (M_{etaeta})^{-1}: conditional covariance Var(eta | y) at current (G,R)
    M_etaeta_inv <- chol2inv(chol(M_etaeta))
    
    ## --- Residuals and trace-adjustment matrices (Algorithm 1) ---
    
    # Conditional residuals
    r_hat <- y - X %*% beta - Z %*% eta
    
    if (!REML) {
      # ML trace-adjustment matrices:
      #   T_tau   = Var(eta | y) = (M_{etaeta})^{-1}
      #   T_sigma = Var(Z eta | y) = Z (M_{etaeta})^{-1} Z'
      T_sigma <- Z %*% M_etaeta_inv %*% t(Z)
      T_tau   <- M_etaeta_inv
      
    } else {
      # REML trace-adjustment matrices:
      #   T_tau   = Var(eta_hat - eta) = C_{eta,eta}
      #   T_sigma = Var(r_hat - eps) = [X Z] C [X Z]'
      T_sigma <- XZ %*% C %*% t(XZ)
      T_tau   <- C_etaeta
    }
    
    ## --- M step updates (tau^2 and sigma^2) ---
    
    # sigma^2 update:
    #   sigma_hat^2 = (r_hat'r_hat)/n + tr(T_sigma)/n
    rss_r        <- as.numeric(t(r_hat) %*% r_hat)
    trace_Tsigma <- sum(diag(T_sigma))
    sigma2_new   <- (rss_r + trace_Tsigma) / n
    
    # tau^2 update:
    #   tau_hat^2 = (eta'eta)/q + tr(T_tau)/q
    q_eta      <- as.numeric(t(eta) %*% eta)
    trace_Ttau <- sum(diag(T_tau))
    tau2_new   <- (q_eta + trace_Ttau) / q
    
    ## --- Convergence check ---
    
    delta <- max(
      abs(sigma2_new - sigma2) / (sigma2 + 1e-8),
      abs(tau2_new   - tau2)   / (tau2   + 1e-8)
    )
    
    sigma2 <- sigma2_new
    tau2   <- tau2_new
    
    if (delta < tol) break
  }
  
  list(
    beta   = drop(beta),
    eta    = drop(eta),
    tau2   = tau2,
    sigma2 = sigma2,
    G      = tau2 * diag(q),
    R      = sigma2 * diag(n),
    iter   = iter,
    REML   = REML,
    
    # --- objects for inspection (final iteration) ---
    M            = M,
    C            = C,
    M_etaeta_inv = M_etaeta_inv,
    C_etaeta     = C_etaeta,
    r_hat        = drop(r_hat),
    T_tau        = T_tau,
    T_sigma      = T_sigma,
    trace_Ttau   = trace_Ttau,
    trace_Tsigma = trace_Tsigma
  )
}
# em_lmm(y, X, Z, REML = TRUE)
\end{lstlisting}

}
\medskip

\section{\texttt{lme4} validation}\label{sec:lme4}

To validate the EM algorithm implementation in Appendix~\ref{app:code}, we compared
its output with estimates from \texttt{lme4}'s \texttt{lmer()} \citep{bates2015}
using a Gaussian one-way random-effects model with three fixed-effect coefficients (an
intercept and two continuous covariates) and a single random-intercept variance
component. For each fit we ran \texttt{em\_lmm()} under both ML and REML and fit
the corresponding model in \texttt{lmer()} with matching settings.
Table~\ref{tab:lme4-validation} reports the true parameter values together with
the EM and \texttt{lme4} estimates of the fixed effects, variance components,
and (RE)ML log-likelihoods; the EM implementation reproduces the \texttt{lmer()}
results to the displayed precision.

A minimal example of extracting compatible $(\bsy,\bsX,\bsZ)$ from an \texttt{lme4}
fit is:
\medskip

{
\begin{lstlisting}
library(lme4)
fit <- lmer(y ~ x1 + x2 + (1 | grp), data = dat, REML = TRUE)
y <- getME(fit, "y")
X <- getME(fit, "X")
Z <- t(getME(fit, "Zt"))   
\end{lstlisting}
}
\medskip

\begin{table}[t]
\centering
\caption{Comparison of \texttt{em\_lmm} (Appendix~\ref{app:code}) and \texttt{lme4} estimates (ML vs.\ REML) for a one-way random-effects model. REML and ML log-likelihood values are on different likelihood scales and are shown only for within-method validation.}

\label{tab:lme4-validation}
\begin{tabular}[t]{llrrrrrr}
\toprule
method & likelihood & $\beta_0$ & $\beta_1$ & $\beta_2$ & $\tau^2$ & $\sigma^2$ & logLik\\
\midrule
Truth &  & 2.0000 & 1.0000 & -0.5000 & 1.0000 & 1.0000 & ---\\
\texttt{em\_lmm} & REML & 1.9684 & 1.3937 & -0.6070 & 0.8616 & 0.6785 & -40.61296\\
\texttt{lme4} & REML & 1.9684 & 1.3937 & -0.6070 & 0.8616 & 0.6785 & -40.61296\\
\texttt{em\_lmm}& ML & 1.9673 & 1.3910 & -0.6026 & 0.7169 & 0.6167 & -39.02332\\
\texttt{lme4} & ML & 1.9673 & 1.3910 & -0.6026 & 0.7169 & 0.6167 & -39.02332\\
\bottomrule
\end{tabular}
\end{table}

\bibliographystyle{plainnat}
\bibliography{tc_reml}

\begin{thebibliography}{25}
\providecommand{\natexlab}[1]{#1}
\providecommand{\url}[1]{\texttt{#1}}
\expandafter\ifx\csname urlstyle\endcsname\relax
  \providecommand{\doi}[1]{doi: #1}\else
  \providecommand{\doi}{doi: \begingroup \urlstyle{rm}\Url}\fi

\bibitem[Bates et~al.(2015)Bates, M{\"a}chler, Bolker, and Walker]{bates2015}
D.~Bates, M.~M{\"a}chler, B.~Bolker, and S.~Walker.
\newblock Fitting linear mixed-effects models using {\texttt{lme4}}.
\newblock \emph{Journal of Statistical Software}, 67\penalty0 (1):\penalty0
  1--48, 2015.

\bibitem[Broatch et~al.(2018)Broatch, Green, and Karl]{broatch2018}
Jennifer Broatch, Jennifer Green, and Andrew Karl.
\newblock {RealVAMS: An R Package for Fitting a Multivariate Value-Added Model
  (VAM)}.
\newblock \emph{The R Journal}, 10\penalty0 (1):\penalty0 22--30, 2018.
\newblock \doi{10.32614/RJ-2018-033}.
\newblock URL \url{https://doi.org/10.32614/RJ-2018-033}.

\bibitem[Dempster et~al.(1977)Dempster, Laird, and Rubin]{dempster1977}
A.~P. Dempster, N.~M. Laird, and D.~B. Rubin.
\newblock Maximum likelihood from incomplete data via the {EM} algorithm.
\newblock \emph{Journal of the Royal Statistical Society, Series B},
  39\penalty0 (1):\penalty0 1--38, 1977.

\bibitem[Diffey et~al.(2017)Diffey, Smith, Welsh, and Cullis]{diffey2017}
S.~M. Diffey, A.~B. Smith, A.~H. Welsh, and B.~R. Cullis.
\newblock A new {REML} (parameter expanded) {EM} algorithm for linear mixed
  models.
\newblock \emph{Australian \& New Zealand Journal of Statistics}, 59\penalty0
  (4):\penalty0 433--448, 2017.

\bibitem[Harville(1974)]{harville1974}
David~A. Harville.
\newblock Bayesian inference for variance components using only error
  contrasts.
\newblock \emph{Biometrika}, 61\penalty0 (2):\penalty0 383--385, 1974.

\bibitem[Harville(1977)]{harville77}
David~A. Harville.
\newblock The use of linear-model methodology to rate high school or college
  football teams.
\newblock \emph{Journal of the American Statistical Association}, 72\penalty0
  (358):\penalty0 278--289, 1977.
\newblock \doi{10.1080/01621459.1977.10480991}.

\bibitem[Henderson et~al.(1959)Henderson, Kempthorne, Searle, and von
  Krosigk]{henderson1959}
C.~R. Henderson, O.~Kempthorne, S.~R. Searle, and C.~M. von Krosigk.
\newblock The estimation of environmental and genetic trends from records
  subject to culling.
\newblock \emph{Biometrics}, 15\penalty0 (2):\penalty0 192--218, 1959.
\newblock \doi{10.2307/2527669}.

\bibitem[Karl et~al.(2024{\natexlab{a}})Karl, Broatch, and Green]{realvams2024}
Andrew Karl, Jennifer Broatch, and Jennifer Green.
\newblock \emph{{RealVAMS: Multivariate VAM Fitting}}, 2024{\natexlab{a}}.
\newblock URL \url{https://CRAN.R-project.org/package=RealVAMS}.
\newblock R package version 0.4-6.

\bibitem[Karl et~al.(2024{\natexlab{b}})Karl, Yang, and Lohr]{gpvam2024}
Andrew Karl, Yan Yang, and Sharon~L. Lohr.
\newblock \emph{{GPvam: Maximum Likelihood Estimation of Multiple Membership
  Mixed Models Used in Value-Added Modeling}}, 2024{\natexlab{b}}.
\newblock URL \url{https://CRAN.R-project.org/package=GPvam}.
\newblock R package version 3.2-0.

\bibitem[Karl(2012)]{karl2012}
Andrew~T. Karl.
\newblock The sensitivity of college football rankings to several modeling
  choices.
\newblock \emph{Journal of Quantitative Analysis in Sports}, 8\penalty0
  (3):\penalty0 1--44, 2012.
\newblock \doi{10.1515/1559-0410.1471}.

\bibitem[Karl and Broatch(2023)]{mvglmmRank2023}
Andrew~T. Karl and Jennifer Broatch.
\newblock \emph{{mvglmmRank: Multivariate Generalized Linear Mixed Models for
  Ranking Sports Teams}}, 2023.
\newblock URL \url{https://CRAN.R-project.org/package=mvglmmRank}.
\newblock R package version 1.2-4.

\bibitem[Karl and Zimmerman(2021)]{karl2021}
Andrew~T. Karl and Dale~L. Zimmerman.
\newblock A diagnostic for bias in linear mixed model estimators induced by
  dependence between the random effects and the corresponding model matrix.
\newblock \emph{Journal of Statistical Planning and Inference}, 211:\penalty0
  107--118, 2021.
\newblock \doi{10.1016/j.jspi.2020.06.004}.

\bibitem[Karl et~al.(2013)Karl, Yang, and Lohr]{karl2013}
Andrew~T. Karl, Yan Yang, and Sharon~L. Lohr.
\newblock Efficient maximum likelihood estimation of multiple membership linear
  mixed models, with an application to educational value-added assessments.
\newblock \emph{Computational Statistics \& Data Analysis}, 59:\penalty0
  13--27, 2013.
\newblock \doi{10.1016/j.csda.2012.10.004}.

\bibitem[Karl et~al.(2014)Karl, Yang, and Lohr]{karl2014}
Andrew~T. Karl, Yan Yang, and Sharon~L. Lohr.
\newblock Computation of maximum likelihood estimates for multiresponse
  generalized linear mixed models with non-nested, correlated random effects.
\newblock \emph{Computational Statistics \& Data Analysis}, 73:\penalty0
  146--162, 2014.
\newblock \doi{10.1016/j.csda.2013.11.019}.

\bibitem[Laird and Ware(1982)]{lairdware1982}
N.~M. Laird and J.~H. Ware.
\newblock Random-effects models for longitudinal data.
\newblock \emph{Biometrics}, 38\penalty0 (4):\penalty0 963--974, 1982.

\bibitem[Laird et~al.(1987)Laird, Lange, and Stram]{lairdlange1987}
N.~M. Laird, N.~Lange, and D.~Stram.
\newblock Maximum likelihood computations with repeated measures: application
  of the {EM} algorithm.
\newblock \emph{Journal of the American Statistical Association}, 82\penalty0
  (397):\penalty0 97--105, 1987.

\bibitem[Lindstrom and Bates(1988)]{lindstrombates1988}
Michael~J. Lindstrom and Douglas~M. Bates.
\newblock {Newton–Raphson} and {EM} algorithms for linear mixed-effects
  models for repeated-measures data.
\newblock \emph{Journal of the American Statistical Association}, 83\penalty0
  (404):\penalty0 1014--1022, 1988.

\bibitem[McCulloch et~al.(2008)McCulloch, Searle, and Neuhaus]{mcculloch2008}
C.~E. McCulloch, S.~R. Searle, and J.~M. Neuhaus.
\newblock \emph{Generalized, Linear, and Mixed Models}.
\newblock Wiley, Hoboken, NJ, 2nd edition, 2008.

\bibitem[McLachlan and Krishnan(2008)]{mclachlankrishnan2008}
Geoffrey~J. McLachlan and Thriyambakam Krishnan.
\newblock \emph{The {EM} Algorithm and Extensions}.
\newblock Wiley, Hoboken, NJ, 2nd edition, 2008.

\bibitem[Meng and Rubin(1993)]{mengrubin1993}
Xiao-Li Meng and Donald~B. Rubin.
\newblock Maximum likelihood estimation via the {ECM} algorithm: A general
  framework.
\newblock \emph{Biometrika}, 80\penalty0 (2):\penalty0 267--278, 1993.
\newblock ISSN 00063444.
\newblock URL \url{http://www.jstor.org/stable/2337198}.

\bibitem[Patterson and Thompson(1971)]{patterson1971}
H.~D. Patterson and R.~Thompson.
\newblock Recovery of inter-block information when block sizes are unequal.
\newblock \emph{Biometrika}, 58\penalty0 (3):\penalty0 545--554, 1971.

\bibitem[Rizopoulos et~al.(2009)Rizopoulos, Verbeke, and Lesaffre]{riz09}
Dimitris Rizopoulos, Geert Verbeke, and Emmanuel Lesaffre.
\newblock Fully exponential laplace approximations for the joint modelling of
  survival and longitudinal data.
\newblock \emph{Journal of the Royal Statistical Society: Series B (Statistical
  Methodology)}, 71\penalty0 (3):\penalty0 637--654, 2009.
\newblock \doi{10.1111/j.1467-9868.2008.00704.x}.

\bibitem[Searle et~al.(1992)Searle, Casella, and McCulloch]{searle1992}
S.~R. Searle, G.~Casella, and C.~E. McCulloch.
\newblock \emph{Variance Components}.
\newblock Wiley, New York, 1992.

\bibitem[Stroup et~al.(2018)Stroup, Milliken, Claassen, and
  Wolfinger]{stroup2018}
Walter~W. Stroup, George~A. Milliken, Elizabeth~A. Claassen, and Russell~D.
  Wolfinger.
\newblock \emph{{SAS for Mixed Models: Introduction and Basic Applications}}.
\newblock SAS Institute Inc., Cary, NC, 2018.

\bibitem[Verbyla(1990)]{verbyla1990}
A.~P. Verbyla.
\newblock A conditional derivation of residual maximum likelihood.
\newblock \emph{Australian Journal of Statistics}, 32\penalty0 (2):\penalty0
  227--230, 1990.

\end{thebibliography}

\end{document}